\documentclass[%
reprint,
nofootinbib,
amsmath,amssymb,
aps
]{revtex4-1}
\usepackage{amsmath,amssymb}
\usepackage{graphicx}
\usepackage{dcolumn}
\usepackage{bm}

\usepackage{natbib}

\usepackage{qcircuit}
\usepackage{dsfont}
\usepackage{color}
\usepackage{hyperref}

\begin{document}
	
	
	\title{Quantum Computing for nonlinear differential equations and turbulence}
	
\author{Felix Tennie}

	\email{f.tennie@imperial.ac.uk}

 	\affiliation{Imperial College London, Department of Aeronautics, Exhibition Road, London SW7 2BX, UK} 
  

\author{Sylvain Laizet}


 	\affiliation{Imperial College London, Department of Aeronautics, Exhibition Road, London SW7 2BX, UK} 

\author{Seth Lloyd}


 	\affiliation{Massachusetts Institute of Technology, Cambridge, MA 02139, USA} 
	
	\author{Luca Magri}
 
 	\email{l.magri@imperial.ac.uk}
	\affiliation{Imperial College London, Department of Aeronautics, Exhibition Road, London SW7 2BX, UK} 

 	\affiliation{The Alan Turing Institute, London NW1 2DB, UK}

	\date{\today}
	
	\begin{abstract}
 A large spectrum of problems in classical physics and engineering, such as turbulence, is governed by nonlinear differential equations, which typically  require high-performance computing to be solved. Over the past decade, however, the growth of classical computing power has slowed down because the miniaturisation of chips has been approaching the atomic scale. This is marking an end to Moore's law, which calls for a new computing paradigm:  Quantum computing is a prime candidate. In this paper, we offer a perspective on the current challenges that need to be overcome in order to use quantum computing for the simulation of nonlinear dynamics. We review and discuss progress in the development of both quantum algorithms for nonlinear equations and quantum hardware. We propose   pairings between quantum algorithms for nonlinear equations and quantum hardware concepts. These avenues open new opportunities  for the simulation of nonlinear systems and turbulence. 

	\end{abstract}
	
	\maketitle

	\section{Introduction}

Nonlinear differential equations and their corresponding dynamics are frequently encountered in various fields such as Physics, Engineering, Chemistry, Biology, Economics, and Weather and Climate Science to name but a few \cite{Jordan2007_nonlinear}. Turbulence, originating from the nonlinear evolution of fluids, is a particular example of a scientific and engineering problem governed by nonlinear partial differential equations (the Navier-Stokes equations). Although the Navier-Stokes equations constitute a broadly accepted mathematical model to describe the motions of a turbulent flow, their solutions can be extremely challenging to obtain because of the chaotic and inherently multi-scale nature of turbulence. The smallest scales impact the largest scales, and small changes to boundary conditions, initial conditions, or grid resolution, for example, can have a dramatic impact on the solution.  The turbulent scales are typically separated by many orders of magnitude. A simulation that captures/resolves all of these scales is called direct numerical simulation (DNS). Except for the simplest of problems, simulating turbulent flows requires high-performance computing (HPC). However, even with today’s state-of-the-art algorithms and petascale systems, DNS is only feasible for a small class of problems, namely those at moderate Reynolds numbers (defined as the ratio of inertial forces to viscous forces) and in simple geometries~\cite[e.g.,][]{moin1998direct}. Consequently, our understanding of turbulence is effectively limited by the performance parameters of currently available computing hardware.
 
Fundamentally, information processing is subject to the laws of physics and therefore a number of fundamental limits on computation exist. For instance, because of the Bekenstein bound, the memory density of physical systems of finite size and energy must itself be finite \cite{Bekenstein1981}. Quantum speed limits \cite{Deffner2017} bound the processing speed and Landauer's principle \cite{Landauer1961} imposes a lower limit on the energy consumption of each irreversible computing step.\footnote{Landauer's limit is given by $k_b T \ln (2)$ per operation, where $k_b$ represents the Boltzmann constant and $T$ the absolute temperature. Current computing hardware is two hundred-fold less efficient \cite{Ho2023}. In  fact, information and communication technologies consume $10\%$ of the worldwide electricity production \cite{Gelenbe2023}.} An `ultimate laptop' \cite{Lloyd2000} of $1\,kg$ of matter confined to $1\,l$ of space would be able to compute at a rate of $5\times10^{50}$ operations per second and store about $10^{31}$ bits. These numbers are far from being attained by current `classical'  silicon-based integrated-circuit computing hardware. However, the exponential growth of classical computers over the past six decades in line with Moore's law is coming to a hold due to a number of practical reasons \cite{IRDS2023}. Miniaturisation of computing chips has reached the scale of a few nanometers and consequently the size of transistors is approaching the atomic scale. On the one hand, this is a fundamental lower bound of integrated circuit technology. On the other hand, quantum effects will become relevant. Despite various mitigation strategies such as parallelisation of computational tasks, a fundamentally different computing paradigm is required to further increase computing power and allow for the simulation of complex systems such as turbulent flows. Here, quantum computing is a prime candidate \cite{Nielson2010}.

Originally conceived in the 1980s~\cite{Feynman1982} to improve simulations of high-dimensional quantum systems that were intractable on classical computers, quantum computing was soon recognised as a powerful tool for solving a range of \textit{linear} classical data processing tasks too, such as Fourier transformation, matrix inversion, singular value decomposition to name but a few. These so-called Quantum Linear Algebra operations theoretically offer exponential resource advantages over their classical counterparts \cite{Biamonte2017} and are at the foundation of Quantum Machine Learning. They also form fundamental building blocks of various subsequently designed quantum algorithms. Among these are a range of quantum algorithms for solving linear differential equation systems (e.g.~\cite{Berry2014,Childs2017,Childs2021highprecision}). Owing to the linear evolution of quantum systems, constructing quantum algorithms for solving nonlinear differential equations has been much harder. Therefore, the question is \\

{\it How can we solve nonlinear equations with quantum computers, which are linear machines?} \\

Over the past five years, a number of conceptually different approaches have been presented. They can broadly be assigned to one of three categories: a) hybrid classical-quantum algorithms (e.g.~\cite{Lubasch2020,Kyriienko2021,Pfeffer2022}), b) mean-field quantum solvers (e.g.~\cite{Lloyd2020,Grossardt2024}), and c) linearisation-based solvers\footnote{In this context, \textit{linearisation} refers to the transformation of a (finite-dimensional) nonlinear dynamical system into a linear system with infinite many degrees of freedom.} (e.g.~\cite{Liu2021,Tennie2024}). In the authors' opinion, nonlinear quantum solvers and their efficient implementation on emerging and future quantum hardware constitute a crucial bottleneck for future applications of quantum computing to turbulence and more generally nonlinear systems. In this paper, we thus review quantum nonlinear solvers and those quantum hardware concepts that, in the authors' opinion, appear most promising for future up-scaling. Following that, we offer our perspective on suitable pairings of quantum algorithms and hardware to achieve a practically relevant quantum advantage in the mid-term future. 

This paper is organised as follows. In Sec.~\ref{sec:InitialConsiderations}, we present preliminary considerations on how to simulate turbulent flows on quantum computers. We  argue that the ability for nonlinear processing of data is a {\it conditio sine qua non}. In  Sec.~\ref{sec:QuantAlgorithms}, we provide an overview and classification of currently developed quantum algorithms for integrating nonlinear system dynamics. In order to apply these algorithms for simulating turbulent flows, one requires quantum hardware which at the time of writing is a fast developing field of research. To understand the interplay between quantum algorithms and hardware, we then review  three currently  promising technologies for building a scalable quantum processor in Sec.~\ref{sec:QuantHardware}. In Sec.~\ref{sec:Questions}, we discuss potential pairings of quantum nonlinear solvers and quantum hardware, and their potential for future numerical simulations. In Sec.~\ref{sec:summary}, we summarise the key insights and offer conclusions.

\section{Initial considerations}\label{sec:InitialConsiderations}

Quantum computers are linear machines. Their ability to process state vectors of exponentially large Hilbert spaces efficiently makes them prime candidates for enhancing all linear tasks in simulations of dynamical systems, such as linear stability of fluids \cite{magri2023linear}. Despite these promises, there are a number of challenges. 

First, at the beginning of a numerical simulation, a quantum state vector must be initialised to encode the initial condition. For instance a data set $\boldsymbol{u}_{ini}  = (u_1,u_2,\ldots,u_d)$ can be encoded in the amplitudes of a quantum state, $|\psi_{ini}\rangle \propto \sum_{i} u_i |i\rangle$\footnote{A quantum register, e.g.~formed by a number of $N$ qubits, has a computational basis of states $\{|i\rangle\}_{i=1}^{2^N}$.}, or in rotation angles of individual qubits, $|\psi_{ini}\rangle=(\cos(\theta_i)|0\rangle + \sin(\theta_i)|1\rangle )^{\otimes_d}$, where $\theta_i = u_i/||\boldsymbol{u}||$. While rotation angle encoding is straightforward and only requires $d$ single-qubit gates, it also requires larger quantum registers  of $N = d$ qubits. Amplitude encoding efficiently utilises the exponential size of Hilbert spaces ($N = \lfloor \log_2 (d)\rfloor$), but might create a severe computational overhead depending on the structure of the input data (compare, e.g., Refs.~\cite{Nielson2010,GonzalezConde2024} for recent developments).

Second, during the computational process, classical data defining the dynamical system, e.g.~coefficients of the underlying differential equations, must be transferred to the quantum computer. Generally, this requires a quantum random access memory (QRAM), i.e.~a mechanism for accessing data based on memory addresses given in form of quantum states. It has recently been argued that \textit{``cheap, asymptotically scalable passive QRAM is unlikely with existing proposals''} \cite{Jaques2023}, and therefore additional research is required to close this gap. In the authors' opinion, QRAM constitutes a fundamental requirement for any type of quantum computing, and therefore does not pose a problem specific to the task of simulating nonlinear dynamics and turbulence.

Third, the integration of nonlinear dynamics requires a nonlinear evolution of the quantum degrees of freedom in which the dynamical variables have been encoded. Because the evolution of quantum systems is unitary and governed by the \textit{linear} Schr\"{o}dinger equation, evolving nonlinearly the quantum degrees of freedom is a fundamental challenge on which we focus our attention in this perspective paper. One particular aspect of the conundrum \textit{linear vs.~nonlinear} is that quantum states are normalised with respect to the $L_2$ norm. However, most sets of dynamically evolving classical data do not meet this condition. Therefore, non-unitary evolution steps are required, which can be achieved by projective measurements and open system dynamics. 

Fourth, the final output of a numerical simulation on a quantum computer is to be encoded in a quantum state. Generally, reading out data, e.g.~determining all quantum amplitudes, requires an exponential overhead which renders the quantum computing approach non-competitive. Hence, it is necessary to identify `macroscopic' quantities, e.g.~averages or higher momenta of sets of dynamical variables, which  can be efficiently sampled from the final quantum state. The recently developed concept of classical shadows \cite{Huang2020} offers a new approach to the task of extracting key information from a quantum computation output.

In summary, the above challenges 1, 2 and 4 are a common aspect of any quantum computation. However, nonlinear processing of data is a challenge that is of particular importance in the simulation of turbulence and nonlinear systems. Therefore, we review key quantum nonlinear solver algorithm concepts in Sec.~\ref{sec:QuantHardware}.

\begin{table*}
	\centering

	\begin{tabular}{p{2cm}p{2.2cm}p{12.1cm}}
		
		\textbf{Developer} & \textbf{Technology} & \textbf{Current status and roadmap}  \\

		\hline
		IBM
		& Superconduc- ting circuits
		& Condor processor (1121 qubits) and the serial Heron processors (133 qubits), which can be connected using classical communication, mark the forefront of development. Future work to be directed at delivering 200 logical error-corrected qubits, capable of executing $10^8$ gates by 2029; further upscaling to 2000 logical qubits in the 2030s \cite{IBM2024}.\\
		\hline
		Google \newline Quantum AI
		& Superconduc- ting circuits
		& Claim of first demonstration of quantum supremacy on the Sycamore processor (54 physical qubits) in 2019 \cite{Arute2019}; current work directed towards a logical qubit capable of executing $10^6$ gates by 2025; future upscaling to achieve $10^6$ physical qubits as a basis for an error-corrected quantum computer within the next decade \cite{GoogleAIQuantum2024}.\\
		\hline
		Atom Computing
		& Neutral Atoms
		& Company has announced the creation of a 1180-qubit processor based on nuclear spins of optically trapped neutral atoms in 2023 \cite{AtomComputing2023}. \\
		\hline
		Rigetti
		& Superconduc- ting circuits
		& Current efforts are focused on improving the fidelity of gates for a 84-physical-qubits processor, followed by the development of the Lyra 336-physical-qubits processor in 2024 \cite{Rigetti2023}.\\
		\hline
		QuEra
		& Neutral Atoms
		& In 2023, demonstration of a programmable quantum processor based on 48 encoded logical qubits operating with up to 280 physical qubits \cite{Bluvstein2023} in reconfigurable neutral-atom arrays. Future plans seek to upscale the processors to more than $10^4$ physical qubits, and encoding more than 100 logical qubits therein \cite{QuEra2023}. \\
		\hline
		Quantinuum
		& Trapped Ions
		& Current H2 processors offers 32 fully connected qubits. Future plans involve achieving 10 logical error-corrected qubits by 2025 with the perspective of upscaling the technology to 1000 logical qubits in the long term \cite{Quantinuum2024}.\\
		\hline
		Pasqal
		&Neutral Atoms
		& In 2024, Pasqal has developed quantum processors with about 100 qubits. By 2026, Pasqal aims to build a processor with 10,000 physical qubits and scalable logical qubits architecture \cite{Pasqal2024}.\\
		\hline
		IonQ
		& Trapped Ions
		& IonQ measures the performance of their quantum processors in terms of Algorithmic Qubits (AQs), which are benchmarked by use-cases in optimization, quantum simulation and quantum machine learning. The company has set out to increase the current value of 32 AQs to 1024 AQs by 2028 \cite{IonQ2024}.\\
		\hline
		%
		
	\end{tabular}
	
	\caption{Illustrative selection of quantum hardware developers, current achievements and future plans that could become significant for future quantum-algorithm-powered simulations of nonlinear dynamics.}\label{tab:roadmaps}
\end{table*}

 \section{Quantum Algorithms for nonlinear dynamics}\label{sec:QuantAlgorithms}
 In this section, we  provide an overview about the different types of quantum algorithms that have already been developed to integrate nonlinear dynamics. These algorithms can mainly be grouped into three categories: hybrid classical-quantum algorithms, mean-field quantum solvers and schemes that apply linearisation techniques to a nonlinear system before integrating the resulting linear equations of motion on quantum computers. The algorithms show considerable differences in key properties and characteristics, such as the chosen method of encoding data, quantum hardware requirements and scaling advantages. Together with prospective future quantum hardware developments, the structural properties of the quantum nonlinear solvers play a vital role in determining the future impact of quantum computing on simulations of turbulence and nonlinearities.

 \subsection{Hybrid Classical-Quantum Algorithms}\label{sec:HybridQuantSolvers}
 
 The lack of scalable error-corrected quantum hardware has motivated various approaches to utilise Noisy Intermediate-Scale Quantum (NISQ) hardware for practical applications. In the context of integrating nonlinear dynamics, all ans\"atze cast differential equations into a minimisation problem whose optimum represents the (approximate) solution. During the computation, a loss function is evaluated for a parametrised trial solution on a quantum processor while the variation and updating of parameters is executed by a classical processor.
 
 In the scheme proposed by Lubasch et al \cite{Lubasch2020}, the loss function is formed by employing Tensor Networks on multiple copies of a quantum state whose amplitudes encode a trial solution. Following a non-local interaction of these copies with an ancilla qubit, the loss function is evaluated by a projective measurement. Although the employed amplitude encoding offers an exponential resource advantage over classical data encoding, and therefore opens the possibility for highly refined meshes, analytic bounds on the convergence of the variational part of this algorithm have still to be developed. The algorithm has been formulated for a gate-based quantum computing framework and requires high connectivity to implement the non-local operations.
 
A different hybrid classical-quantum framework was introduced by Kyriienko et al \cite{Kyriienko2021}. Based on a quantum feature map which encodes data in form of rotation angles of individual qubits, a finite set of Chebyshev polynomials is used to fit and approximate a solution function. As in Ref.~\cite{Lubasch2020}, the loss function is cast into a gate-based quantum circuit. However, the circuit does not comprise computationally expensive many-qubit gates and is therefore suitable for hardware architectures with limited connectivity. The overall computational cost can however not yet be assessed since analytic bounds on the convergence of the optimisation process have not been established.

A third  hybrid ansatz is motivated by classical data-driven reservoir computing models, a type of recurrent neural network. Quantum reservoir computing \cite{Pfeffer2022,Oz2023} utilises the exponential size of Hilbert spaces to form a large dimensional parametrised reservoir state that replaces the classical reservoir state. Using real or synthetic data, the network, represented by a parametrised quantum circuit, is trained and subsequently used to predict new time series data. Quantum reservoir computing is a heuristic method that has successfully been demonstrated to capture essential features of low dimensional non-linear systems such as Lorentz 63. Although it has limited quantum hardware connectivity demands, training requires frequent sampling of the circuit outputs which renders the scheme computationally expensive. Future work must yet establish analytic bounds on these quantities.

\subsection{Mean-field nonlinear solvers}\label{sec:MeanFieldQuantSolvers}

The mean-field nonlinear quantum algorithm proposed in Ref.~\cite{Lloyd2020} utilises symmetric interactions between $n\gg 1$  identical copies of a quantum state $|\psi\rangle = \sum_i u_i |i\rangle$ to generate an approximate nonlinear evolution of an individual copy. The dynamical variables are encoded in the amplitudes $u_i(t)$. In its original form, the mean-field nonlinear solver uses a matrix inversion scheme to implement a set of Forward-Euler steps which is solved by a Quantum Linear System Algorithm (QLSA) such as  the Harrow–Hassidim–Lloyd (HHL) algorithm  \cite{Harrow2009}. The output is a large history state formed by multi-copy states at different discrete time steps. Since quantum matrix inversion requires extensive quantum hardware resources, one can alternatively employ a quantum matrix multiplication followed by measurement-and-postselection of an ancilla qubit. This can be well illustrated on the example of the one-dimensional system $\dot{x}(t) = x - x^3$: First, the quantum register of $n$ qubits must be initialised in the state $|\Psi_{t_0}\rangle = (x_0|0\rangle + \sqrt{1-x_0^2}|1\rangle)^{\otimes^n}$, where $x_0 = x(t_0)$ is the initial condition. Then Forward-Euler steps of stepping size $\Delta t$ are implemented by quantum matrix multiplications, $|\Psi_{t_{i+1}}\rangle = (1 + \Delta t A) |\Psi_{t_{i}}\rangle$, where $A = \sum_{j=1}^{n} |0_j\rangle \langle 0_j| + 2/n^2\sum_{j\neq k}^{n} |0_j\rangle \langle 0_j |\otimes |0_k\rangle \langle 0_k|$. This can either be achieved by unitary block encoding \cite{Lloyd2021}, or alternatively by a linear-addition-of-unitaries scheme recently developed in Ref.~\cite{Tennie2024}. The latter requires two ancilla qubits and a controlled Hamiltonian simulation of $H = i(1 + \Delta t A)\otimes |0\rangle\langle 1| + h.c.$\footnote{Here, $h.c.$ refers to the hermitian conjugate of the preceding operator.}. Because of the mean-field character of this time stepping, the states $|\Psi_{t_{i+1}}\rangle$ approximately remain product states of propagated individual copies: $|\Psi_{t_k}\rangle \approx (x_k|0\rangle + \sqrt{1-x_k^2}|1\rangle)^{\otimes^n}$. By measuring individual copies, one can then retrieve the output $x(t_k) = x_k$ of the numerical integration.

For arbitrary nonlinear systems, any of the above mentioned implementations of the mean-field solver generally requires Hamiltonian simulation of permutation-symmetric Hamiltonians
\begin{equation}\label{eq:MeanFieldHamiltonian}
	H = \sum_{\varsigma \in S_n} H_{\varsigma(1),\ldots,\varsigma(m)},
\end{equation}
where $S_n$ denotes the group of permutations and $m$ is the leading degree of the polynomial $f$ in the underlying differential equation, $\dot{\boldsymbol{x}}  = \boldsymbol{f}(\boldsymbol{x})$.\footnote{It should be noted that the mean-field solver naturally generates polynomials with odd-order terms. Even order terms can be implemented by using dummy variables which are set to 1 and kept constant throughout the numerical integration, e.g. $x^2 =  \bar{x} x^2$ with $\bar{x} = 1$.} Figuratively, this can viewed as a requirement for all copies to `shake hands'. As a consequence, suitable quantum hardware should offer a large connectivity between the individual copies to reduce the computational overhead required for swapping pairs of copies, which is further discussed in Sec.~\ref{sec:Questions}. For more details on the mean-field solver algorithm and the above example, the reader is referred to  Ref.~\cite{Lloyd2020}.

Analytic bounds on the performance of the mean-field nonlinear quantum solver have been established: the number of copies $n$ grows linearly with the integration time. Thanks to amplitude encoding, this scheme also offers an exponential resource gain with respect to the number of grid points.  The structure of the Hamiltonian \eqref{eq:MeanFieldHamiltonian}, however, implies that not all quantum hardware architectures are equally suitable, and those with an anywhere-to-anywhere connectivity are advantageous.

\subsection{Linearisation-based Quantum Algorithms}\label{sec:LinearisationQuantSolvers}

Since quantum computers are effectively linear machines, it seems plausible to consider whether a faithful linearisation of a nonlinear system can be used as a precursor for integration. So far, two different methods have been proposed.

In Ref.~\cite{Liu2021}, Carlemann linearisation was employed to tackle a range of nonlinear systems in which dissipation dominates the combined strength of nonlinearities and inhomogeneities.  For any finite dimensional nonlinear differential equation system, Carlemann linearisation yields an infinite set of linear differential equations which is truncated and subsequently solved by a QLSA. This approach has been well studied and analytic bounds on its performance are established. They show that it offers a quantum advantage on scalable future quantum computers. However, because of the considerable quantum hardware requirements for running a QLSA (both number of qubits and circuit depth), NISQ hardware may not be sufficient to support it. 

Another linearisation approach is based on the Fokker-Planck equation \cite{Tennie2024}. The Fokker-Planck equation describes the evolution of the distribution function of dynamical variables $\rho(\boldsymbol{x},t)$ subject to a generally nonlinear drift $\boldsymbol{f}(\boldsymbol{x})$ and diffusion $D$: $\partial_t \rho =  \sum_{x_i} \partial_{x_i} (\partial_{x_i}( D \rho)-\rho)$. It is a linear partial differential equation which can be spatially semi-discretised and then integrated. Using amplitude encoding for the values of a discretised distribution function, large grids of small mesh size can be efficiently stored in a quantum register. The integration over discretised time steps is carried out by repeated Hamiltonian simulation and measurement-and-postselection steps. For instance, for the above one-dimensional example $\dot{x}(t) = x - x^3$, semi-discretisation of the spatial domain on an interval ${x_i} \in [-L,L]$ results in a Master equation of the form $\dot{\boldsymbol{p}}= \boldsymbol{F} \boldsymbol{p}$, where $p_i$ is the probability of finding the dynamical variable at a value $x_i\pm \Delta x$ and $\boldsymbol{F}$ only contains diagonal and first-off diagonal elements. Similarly to the mean-field solver, Forward-Euler steps can be implemented by unitary block encoding or by linear addition of unitaries. The latter requires Hamiltonian simulation of $H = i(1 + \Delta t \boldsymbol{F})\otimes |0\rangle\langle 1| + h.c.$. Note that this embedding is necessary since $\boldsymbol{F}$ is a non-normal operator and can therefore not directly be simulated using Hamiltonian evolution.

For arbitrary nonlinear systems, because of the structure of the Fokker-Planck equation, the Hamiltonians required for this procedure are sparse and exhibit a native tensor product structure beneficial to quantum simulation. Unlike other quantum algorithms for solving nonlinear dynamics, the output of this scheme is a distribution function and not a set of dynamical variables. It therefore is suitable for modelling that requires ensemble forecasts and a notion of the propagation of uncertainty in initial values.

This concludes our overview on currently developed quantum `software' for integrating nonlinear dynamics. In order to assess the ability of the above quantum algorithms to compete with classical state-of-the-art numerical simulations of turbulence and nonlinear systems, we review potential quantum hardware candidates in Sec.~\ref{sec:QuantHardware}.

\section{Quantum Computing Hardware}\label{sec:QuantHardware}

\begin{table*}
	\centering

	\begin{tabular}{|p{4cm}|p{4cm}|p{3cm}|p{3cm}|}
		\hline
		\textbf{} & \textbf{Superconducting QC} & \textbf{Trapped-Ion QC} & \textbf{Neutral Atom QC} \\

		\hline
		Qubit lifetime
		& $\sim$ 100 $\mu$s
		& $\sim$ several minutes 
            & $\sim$ several seconds \\
            \hline
		
		Single-qubit and two-qubit gate fidelity $\mathcal{F}$
		& 0.9999, 0.99
		& 0.999999, 0.998 
            & 0.996–0.999, 0.955–0.995 \\
            \hline
		
		Gate execution time
		& $\sim$ $10-100$ns
		& $\sim$ $10-100\mu$s
            & 400ns - 2$\mu$s \\
            \hline
	
		Connectivity
		& 4:1
		& 40:1 
            & 10:1 - 20:1 \\
            \hline
	
		Number of physical qubits
		& $\sim$ 1000
		& $\sim$ 40 
            & $\sim$ 1000 \\
            \hline

	\end{tabular}
	
	\caption{Typical performance parameters of current quantum computing hardware \cite{Wintersperger2023,Bruzewicz,Krantz2019}.}\label{tab:TechParameters}
\end{table*}

Quantum algorithms require a physical realisation of qubits and unitary evolution, i.e.~quantum hardware, to be executed. For more than two decades, the quest for scalable quantum hardware that is robust against various types of environmental noise has generated a number of different concepts. In the authors' opinion,  three currently promising candidates are {\it trapped-ion, superconducting circuit and neutral atom quantum processors}. For some other technological concepts, such as photonic quantum computing or semiconductor spin qubits, at the time of writing this article only quantum processors with a few qubits with limited gate fidelities have been developed and, thus, it will  not  be further discussed.

In this section, we  briefly review the main properties of trapped-ion, superconducting circuit and neutral atom quantum processors, in particular those properties relevant to the quantum algorithms discussed in Sec.~\ref{sec:QuantAlgorithms}. In Tab.~\ref{tab:roadmaps}, we show a selection of quantum hardware developers and the current status and future roadmaps of their products. In Tab.~\ref{tab:TechParameters} we list typical performance parameters of current quantum processors.

Generally controlled by external classical fields, quantum hardware can be operated to provide commonly known gates of a universal gate set, such as single qubit rotations and the CNOT gate. In addition to this, it can also directly be used for Hamiltonian simulation of Hamiltonians that are structurally similar to the native Hamiltonian of the quantum hardware. The suitable choice of one of these two modes of operation depends on the specific quantum algorithm to be executed.

\subsection{Trapped-Ion Quantum Computing}\label{sec:TrappedIonQC}

\begin{figure*}
 	\includegraphics[width=0.5\linewidth]{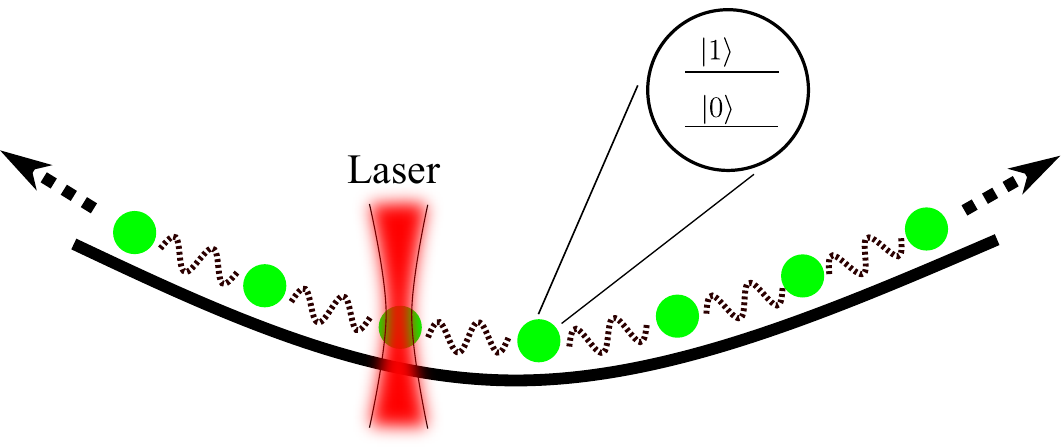}\hspace{1cm}	\includegraphics[width=0.3\linewidth]{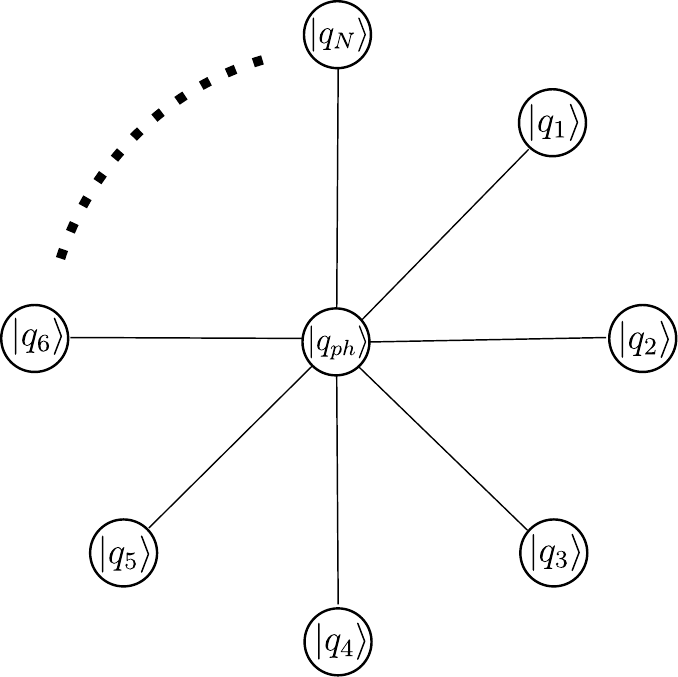}
 	\caption{Left: schematic display of a trapped-ion quantum computer. The ions (green spheres) are trapped within a harmonic potential and are subject to Coulomb forces (dashed springs). Laser pulses can be used to drive transitions between the internal degrees of freedom that represent individual qubits. Laser pulses can also be used to couple internal and external (phononic) degrees of freedom. The resulting connectivity graph is shown on the right: the star-like connectivity between the internal state qubits $|q_i\rangle$ and the phononic qubit $|q_{ph}\rangle$ can be used to effectively mediate interactions between arbitrary pairs of qubits leading to an effective anywhere-to-anywhere connectivity of trapped-ion quantum computers.}\label{fig:TrappedIonQC}
\end{figure*}

 Trapped-ion quantum computing \cite{Bruzewicz}  utilises internal degrees of freedom of ionised atoms trapped by time-varying electric fields (cf.~Fig.~\ref{fig:TrappedIonQC}). Both lasers and microwave radiation can be used to drive transitions between the internal states, and to entangle the ions' internal degrees of freedom to external motional degrees of freedom in a chain of ions, known as phonons. The latter corresponds to neighbour-neighbour interactions on a star-like lattice where all outer sites are linked with a central site that represents the phononic degree of freedom. It provides trapped-ion quantum computers with an effective anywhere-to-anywhere connectivity. For $N$ ions, the native Hamiltonian can conceptually be written as
 \begin{equation}\label{eq:TrappedIonNativeHamiltonia}
 	H = \sum_{i=1}^{N} H_i(t) + \lambda_i(t)(\sigma_i^+ a + \sigma_i^- a^\dagger).
 \end{equation}
 The single qubit Hamiltonians $H_i$ allow for arbitrary single qubit operations and can be implemented with high fidelity\footnote{The fidelity $\mathcal{F}$ \cite{Nielson2010} provides a measure for the overlap of the actual with the desired outcome of a quantum evolution.} ($\mathcal{F}\geq 1 - 10^{-5}$). 
 The operators $\sigma_i^{\pm}$ transfer the $i$-th qubit from the ground to the excited state and vice versa, whereas the operators $a$ and $a^\dagger$ create and destroy phonons.
 The interaction between the ions' internal degrees of freedom and the phonon mode can be used to construct effective interactions between any pair of qubits leading to two-qubit gate fidelities of order $\mathcal{F}\geq 1 - 10^{-4}$, 
 which currently exceeds the performance of all competing hardware concepts (cf.~also Tab.~\ref{tab:TechParameters}).
 
 Trapped-ion quantum processors offer qubit coherence times of up to several minutes. With gate execution times of the order of a few dozen microseconds, several thousands of gates can in principle be implemented. Although this is an advantage of trapped-ion quantum processors, their up-scaling to thousands of qubits remains a challenge since it is not feasible to trap more than a couple of dozen ions in a single trapping potential. Ans\"{a}tze for concatenating multiple traps and `shuttelling' ions between different traps are currently explored \cite{Bruzewicz} which might provide a path towards scalable trapped-ion architectures.
 
 \subsection{Superconducting Quantum Computing}

\begin{figure*}
	\centering
	\includegraphics[width=0.9\linewidth]{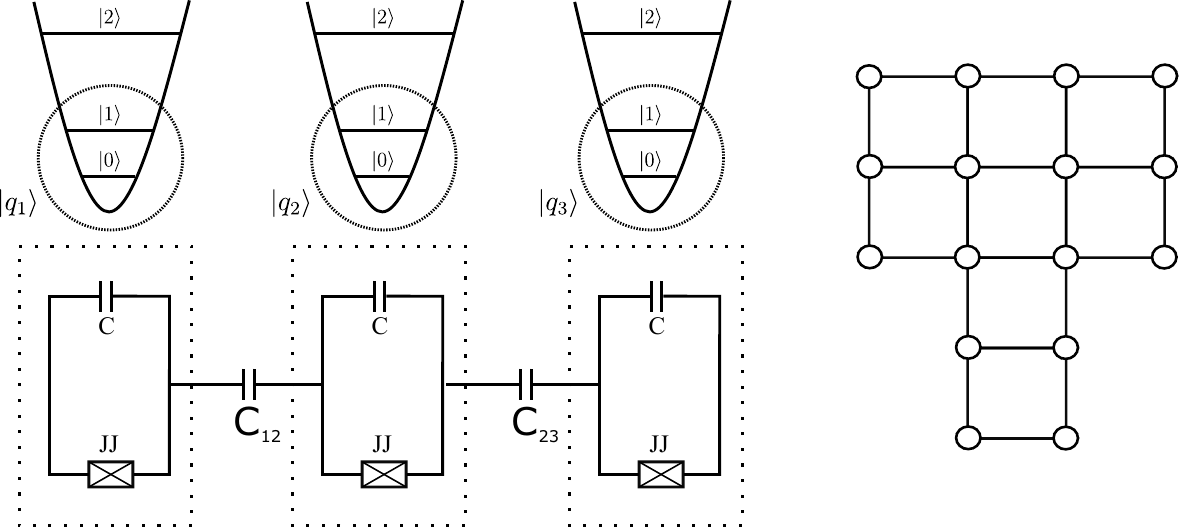}
	\caption{Left: schematic circuit layout of three superconducting qubits on a typical two-dimensional waver. Each qubit $q_i$ is formed by the two lower levels of an anharmonic oscillator comprising of a capacitor $C$ and a nonlinear inductor (Josephson junction $JJ$). The capacitive coupling $C_{ij}$ between pairs of qubits allows for implementation of two-qubit gates. The requirements for this physical link limits the connectivity of this type of quantum processor. Right: a typical connectivity graph of a superconducting circuit quantum processor.}
	\label{fig:supeconductingcircuitquantumcomputer}
\end{figure*}

 Superconducting quantum processors utilise quantum degrees of freedom in nonlinear  $LC$-circuits formed by lithographically manufactured superconducting circuit elements. While various concepts exist that allow for engineering the circuit parameters, the elementary quantum systems can always be described as anharmonic oscillators, c.f.~Fig.~\ref{fig:supeconductingcircuitquantumcomputer}. Capacitive and inductive coupling between pairs of anharmonic oscillators and coupling of individual oscillators to linear resonator circuits leads to effective Hamiltonians of the form \cite{Krantz2019}
 \begin{align}\label{eq:SuperconductingQubitNativeHamiltonian}
 	H = \sum_{i=1}^{N} \big(\omega_i b_i^\dagger b_i &(1 + \alpha_i  b_i^\dagger b_i)
 	+ (\epsilon_i b_i + h.c.)\big)\nonumber \\ 
 	&+ \sum_{\langle i, j \rangle} J_{ij}(b_i b_j^\dagger + h.c.),
 \end{align}
where the operators $b_i$ are ladder operators of the anharmonic oscillators, $\epsilon_i$ are external tunable (classical) microwave amplitudes and $\omega_i$, $\alpha_i$ and $J_{ij}$ are engineerable circuit parameters. The double sum only involves neighbouring pairs of anharmonic oscillators, which by contrast to trapped-ion and neutral atom quantum processors imposes a strong limitation on the connectivity of superconducting quantum processors (cf.~connectivity graph in Fig.~\ref{fig:supeconductingcircuitquantumcomputer}). 

In practise, the two lowest energy eigenstates of each anharmonic oscillator are used to form well-defined qubits. Unlike their `natural' counterparts in ions and neutral atoms, these qubits are not perfectly identical because of  manufacturing imperfections, which remains a source of errors. 
Currently, superconducting qubits have coherence times in the range of hundreds of microseconds and gate execution times in the range of tens to hundreds of nanoseconds. Although this in principle permits circuits depths of several thousand gates, two-qubit gate fidelities currently limited to $\mathcal{F}\approx 0.99$ impose major restrictions on the practical utility of superconducting processors. The limited connectivity and low gate fidelities imply that superconducting quantum processors are not yet suitable for quantum nonlinear solvers described in Sec.~\ref{sec:MeanFieldQuantSolvers} and \ref{sec:LinearisationQuantSolvers}. However, the prospect of scaling superconducting quantum processors to thousands of qubits in the near term future, and continuing work to improve connectivity and gate fidelities makes them interesting candidates for hybrid quantum algorithms as described in Sec.~\ref{sec:HybridQuantSolvers}.

  \subsection{Neutral Atom Quantum Computing}
  
  \begin{figure}
  	\centering
  	\includegraphics[width=0.9\linewidth]{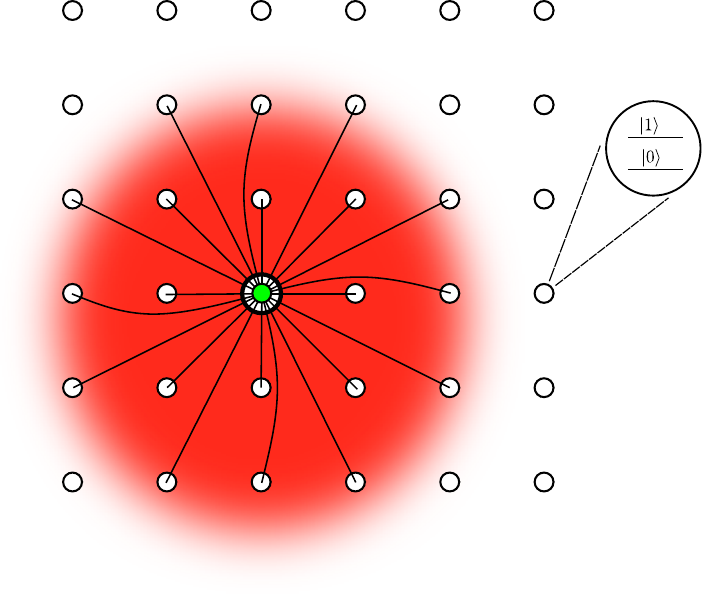}
  	\caption{Schematic depiction of a single two-dimensional layer of optically trapped neutral atoms.  Internal degrees of freedom define individual qubits. An atom excited to a Rydberg state (green) exhibits a strong long-range interaction with other atoms in the optical lattice by the mechanism of the Rydberg blockade. This can be used to achieve a high connectivity of the quantum processor as indicated by the black connecting lines between the excited atom and other atoms.}
  	\label{fig:neutralatomquantumcomputer}
  \end{figure}

Neutral atom quantum processors are based on quantum degrees of freedom in the electronic structure of neutral atoms that are cooled by laser fields and trapped in optical lattices in a vacuum chamber \cite{Henriet2020,Wintersperger2023}. Owing to  their long lifetimes, hyperfine states are typically used to define individual qubits. The internal state of the qubits is manipulated by focused laser beams of tunable frequency and intensity. Interactions between different neutral atom qubits are implemented by the mechanism of the Rydberg blockade: a laser field is used to excite the first qubit into a high-energy Rydberg state which possesses a large dipole moment. This, in turn, suppresses the formation of a Rydberg state of the second qubit when subjected to a laser pulse. The Rydberg blockade is a highly nonlocal interaction which allows for entangling pairs of qubits across many lattice spacings (cf.~Fig.~\ref{fig:neutralatomquantumcomputer}). The effective Hamiltonian may be written as
\begin{align}
	H = \sum_{i=1}^N \Omega_i \sigma^x_i + \delta_i\sigma^x_i + \sum_{i\neq j} \frac{C_3}{r_{ij}^3}(\sigma^x_i \sigma^x_j + (\sigma^y_i \sigma^y_j ),
\end{align}
where $\sigma^{x,y,z}_i$ represent the Pauli matrices associated with the $i^{th}$ qubit. The Raby frequency $\Omega$ depends on the laser intensity, and $\delta$ represents the detuning between the qubit and laser frequency. The dipole-dipole coupling constant $C_3$ and the distance between a pair of neutral atoms $r_{ij}$ determine the two-qubit interaction strength. 

Because of the strong Rydberg interaction, neutral atom quantum processors offer a large connectivity. In addition, it is - in principle - possible to move individual atoms across the lattice and thereby achieve an anywhere-to-anywhere connectivity. 

Neutral atom quantum processors have coherence times of the order of several seconds \cite{Wintersperger2023}. With gate execution times of the order of a microsecond this again permits large circuits depths. Whereas the fidelity of two-qubit gates between physical qubits is currently limited by $\mathcal{F}\approx 0.995$, the remarkable achievement of successful error correction on logical qubits \cite{Bluvstein2023} opens a new path towards scalable quantum computing. A drawback, however, is the long initial quantum state preparation time of several hundreds of milliseconds, which substantially increases the cost for sampling the final quantum state output. Despite this, the absence of manufacturing imperfections (atoms are perfectly identical) and the technological scalability of the number of atoms to be trapped in optical lattices\footnote{The company QuEra has already achieved hundreds of physical qubits and 48 logical error corrected qubits. Future plans seek to achieve 100 logical qubits within the next three years of writing this article (also cf.~Tab.~\ref{tab:roadmaps})} make neutral atom quantum computing a prime candidate for achieving practical utility of the above mentioned quantum algorithms for nonlinear systems. 
 
 \section{Marrying software and hardware}\label{sec:Questions}
\begin{figure*}
    \centering
    \includegraphics[width=0.9\linewidth]{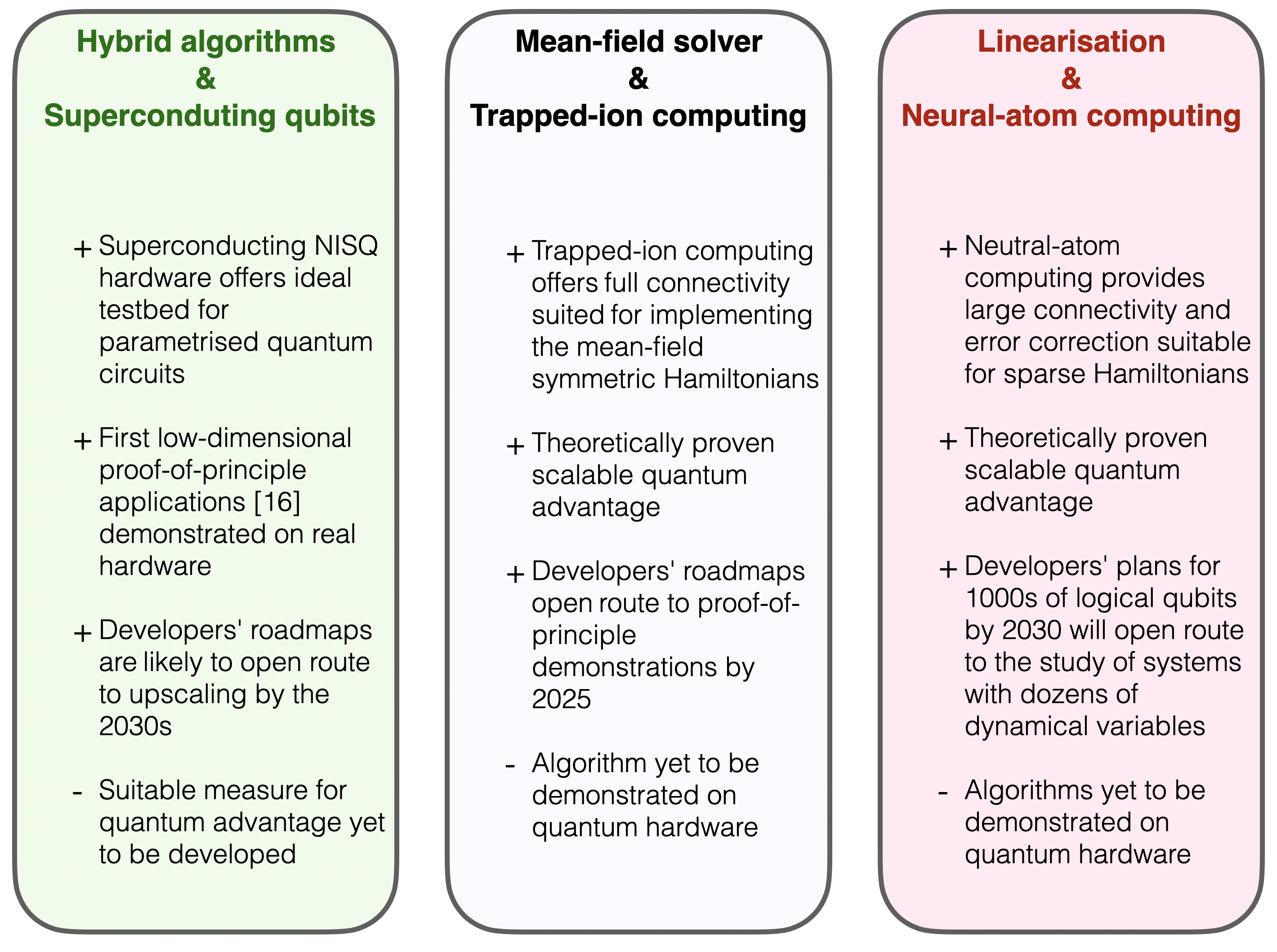}
    \caption{Proposed pairings of quantum nonlinear solvers with different types of quantum hardware concepts.  }\label{fig:suitableMatches}
\end{figure*}

 In this section, we discuss potential 'pairings' of quantum nonlinear solver software (Sec.~\ref{sec:QuantAlgorithms}) and hardware (Sec.~\ref{sec:QuantHardware}). Although most quantum hardware concepts, particularly those presented in Sec.~\ref{sec:QuantHardware}, implement a universal quantum computing model\footnote{Like a classical Turing machine, a universal quantum computer, also know as a Quantum Turing machine, can be used to emulate any quantum evolution within desired precision requiring only a polynomial computational resource overhead. From a complexity point of view, it is therefore equivalent to any other quantum computer. Native operations on gate-based universal quantum computers, for instance, commonly comprise single-qubit rotations and an entangling two-qubit gate and can arbitrarily approximate any unitary evolution.}, because of limited circuit depths, a task-optimised approach is required for successfully running simulations of nonlinear dynamics on quantum hardware in the near to mid-term future. This is analogous to choosing task-optimised classical processors such as Graphical Processing Units or Tensor Processing Units over Central Processing Units. It seems plausible that the future development of quantum processors will become task oriented. In fact, the authors believe that the benefit of current quantum algorithm design is to inform future quantum hardware design. From this perspective, we  analyse future quantum hardware requirements. Having introduced three classes of quantum nonlinear solvers in Sec.~\ref{sec:QuantAlgorithms}, we  argue that the following pairings could bear fruit in the near- to mid-term future: a) Hybrid quantum nonlinear solvers and circuits of superconducting qubits, b) Quantum mean-field solvers and trapped-ion quantum computers, and c) Linearisation-based quantum solvers and neutral atom computing. These pairings (cf.~Fig.~\ref{fig:suitableMatches}) should not be considered as a strict rule; however, there are several advantages in each of these pairings as explained in the next sections. 

 \subsection{Hybrid Classical-Quantum Algorithms and Circuits of Superconducting Qubits}

 Hybrid classical-quantum algorithms commonly require a quantum processor for evaluating a loss function encoded in a parameterised quantum circuit. Often, they have been specifically designed for (noisy) gate-based superconducting quantum processors with limited connectivity (cf.~Fig.~\ref{fig:supeconductingcircuitquantumcomputer}). Indeed, a number of proof-of-principle simulations of nonlinear dynamics have been carried out on quantum processors such as those offered by IBM Quantum (e.g.~\cite{Lubasch2020,Pfeffer2022,Osama2024}), thereby highlighting the prospect of upscaling these algorithmic concepts to problems involving tens of thousands or even millions of degrees of freedom. Since to the best of the authors' knowledge no analytical performance bounds have been derived, and since the notion of a quantum advantage of these hybrid algorithmic concepts must yet be developed, progress in this branch of nonlinear quantum solvers requires a `proof-by-demonstration' upscaling approach. 
 
 Circuits of (noisy) superconducting qubits have over the past few years been upscaled from dozens to hundreds of qubits, and the roadmaps of some developers (cf.~Tab.~\ref{tab:roadmaps}) indicate that more than a million physical qubits might be available around 2030s. While these numbers do not equate to fault-tolerant logical qubits, the corresponding quantum processors can serve as a practical test bed for applying hybrid algorithmic concepts to large-scale problems. For instance, approaches using angular encoding (e.g.~Quantum Reservoir Computing) could be put to a test by running simulations of dynamical systems with a million degrees of freedom, and benchmarking the quantum output with simulations on classical computers. Note that although two-qubit gates in circuits of superconducting qubits only have comparatively moderate fidelities, studies have revealed a robustness of the computational output of minimisation problems \cite{Gentini2020}, and therefore one can conjecture robustness of the hybrid nonlinear quantum solvers to circuit imperfections.

 The `proof-by-demonstration' upscaling approach appears less feasible on trapped-ion and neutral atom quantum computer processors. For trapped-ion quantum computers, current developer roadmaps `only' aim for several thousands of qubits. Albeit these systems are likely to exhibit lower levels of noise and better connectivity in comparison to their superconducting counterparts, both features appear not to play a key role in upscaling the hybrid algorithmic concepts. On neutral atom computing platforms, during each parameter updating step, the measurement of the quantum register required for the evaluation of the loss function produces a practically relevant computational overhead because of subsequent long machine initialisation times (about $0.5$ seconds \cite{Wintersperger2023}). 
 Therefore, hybrid classical-quantum nonlinear solvers are best matched with  superconducting qubit processors.

\subsection{Mean-field solvers and Trapped-Ion Quantum Computers}

Mean-field quantum nonlinear solvers are based on symmetric interactions between a number $n$ of identical quantum state vector copies (cf.~Eq.~\ref{eq:MeanFieldHamiltonian}).
This fundamental structural requirement is best met by a quantum processor with a large connectivity. The two-dimensional architectures of superconducting qubit processors only provide a limited neighbour-neighbour connectivity \eqref{eq:SuperconductingQubitNativeHamiltonian} as illustrated by the example shown in Fig.~\ref{fig:supeconductingcircuitquantumcomputer}. Thus, when implementing a mean-field nonlinear solver in a circuit of superconducting qubits, swapping entire sets of qubits becomes necessary. Owing to the two-dimensional structure of the circuit, the number of swapping operations grows quadratically with $n$. Since $n$ scales linearly with the number of integration steps, the computational overhead  also grows quadratically with the number of integration steps. We conclude that superconducting qubit processors do not offer a suitable platform for mean-field nonlinear solvers.

The native Hamiltonian of trapped-ion quantum processors \eqref{eq:TrappedIonNativeHamiltonia} has the connectivity of a star-like graph, Fig.~\ref{fig:TrappedIonQC}. This can be used to generate an effective second order interaction giving rise to an anywhere-to-anywhere connectivity between sets of qubits. Together with the high fidelity of (two-)qubit operations this makes trapped-ion processors a very suitable platform for mean-field nonlinear solvers. Currently, the upscaling of trapped-ion architectures remains a challenge (cf.~Ref.~\cite{Mouradian2023} and references therein), and the roadmaps of several developers set the goal of more than a thousand logical qubits within the next decade. This would allow for the integration of a nonlinear system of about thousand dynamical variables, for example, a two-dimensional  turbulent  flow~\cite[e.g.,][]{Racca2023}, over one hundred discretised integration steps.\footnote{One thousand dynamical variables can be stored in the quantum amplitudes of $10$ qubits. Integration over $T =100$ steps requires approximately (depending on the strength of the nonlinearity) $n=100$ copies of the quantum state vector, leading to a total number of required qubits of $10\times n = 1000$.}\label{footnoteOnSystemSizes}

Neutral atom computing processors also offer an extensive connectivity, which, however is fundamentally limited by the range of the Rydberg blockade mechanism (cf.~Fig.~\ref{fig:neutralatomquantumcomputer}). In order to implement symmetric interactions, techniques such as shutteling atoms between different lattice sites, are required. Future roadmaps aim for thousands of logical qubits allowing for the integration of systems of sizes similar to the trapped-ion implementation.

Although both platforms, neutral atom and trapped-ion quantum computing, offer similar prospects in the near- to mid-term future, the fact, however, that the connectivity of neutral-atom processors is physically limited, appears to make trapped-ion quantum processors the better long-term candidate for implementing mean-field nonlinear quantum solvers. 

\subsection{Linearisation-based solvers and Neutral Atom Quantum Computers}
 
 Linearisation-based solvers, operated either by employing QLSAs or direct quantum matrix multiplication techniques, rely on efficient Hamiltonian simulation of sparse matrices. 
 
 In the case of the Fokker-Planck framework \cite{Tennie2024}, the Hamiltonians that must be simulated have a specific tensor product structure that can be mapped onto a partition of the quantum register. On a qubit level, the Hamiltonians comprise nonlocal interaction terms, and therefore suitable quantum hardware should offer a large connectivity, making trapped-ion or neutral atom quantum processors suitable candidates. Since the expected circuit depths exceed thousands of layers, platforms must be fault tolerant. 

 In the case of Carlemann linearisation solvers, it has recently been argued \cite{Sanavio2023} that Lattice Boltzmann models of turbulent flows with moderate Reynolds numbers of $10-100$ could possibly be simulated on quantum processors permitting circuit depths of tens of thousands of gates. These requirements may be matched by some of the neutral atom computer developers' roadmaps within the next decade (Tab.~\ref{tab:roadmaps}).

 We conclude that, albeit circuits of superconducting qubits and trapped-ion quantum processors might also become suitable platforms for linearisation-based nonlinear quantum solvers in the mid- to long term future, recent progress in the development of fault-tolerant neutral atom quantum processors \cite{Bluvstein2023} indicates that neutral atom quantum processors are likely to be the first platform to meet the resource requirements of linearisation-based solvers to be applied to a range of physically relevant problems.

 \section{Summary and Conclusion}\label{sec:summary}

We have offered a perspective on the possibility of quantum computers becoming useful tools for simulating nonlinear systems such as turbulent flows. At first, we have outlined four key parts of such a simulation and identified the nonlinear processing step as a bottleneck because of the fundamentally linear nature of quantum computers. We have then discussed three structurally different classes of currently known algorithm concepts for nonlinear quantum solvers. Following that, we have looked at three currently promising quantum hardware technology concepts: circuits of superconducting qubits, trapped-ion and neutral atom quantum processors. By comparing the structural properties and resource requirements of the quantum nonlinear solvers with the parameters of near- to mid-term quantum hardware, we have proposed three pairings which are likely to offer a quantum simulation tool for systems of thousands of nonlinear dynamical variables within the next decade. More research is required to adapt quantum algorithms and quantum hardware further to achieve a practical quantum advantage in simulating nonlinear dynamics within the next one or two decades. 
 
\section{Acknowledgements}
Funding: The authors acknowledge financial support from the UKRI New Horizon grant EP/X017249/1. L.M. is grateful for the support from the ERC Starting Grant PhyCo 949388.

\bibliographystyle{unsrtnat}
\bibliography{references_ImpactOfQuantumComputingOnCFD_Perspective}
		
\end{document}